\documentstyle[aps,pra,preprint]{revtex}
\begin{document}
%
\title{Random-phase approximation study of collective excitations in the
    Bose-Fermi mixed condensate of alkali-metal gases}
\author{T. Sogo, T. Miyakawa, T. Suzuki and H. Yabu}
\address{Department of Physics, Tokyo Metropolitan University, 
         1-1 Minami-Ohsawa, Hachioji, Tokyo 192-0397, Japan}
\date{\today}
\maketitle
\begin{abstract}
    We perform  Random Phase Approximation (RPA) study of collective 
    excitations in the  bose-fermi mixed degenerate gas of Alkali-metal 
    atoms at $T=0$. The calculation is done by 
    diagonalization in a model space composed of particle-hole type 
    excitations from the ground state, the latter being obtained from the 
    coupled Gross-Pitaevskii and Thomas-Fermi equations. We investigate 
    strength distributions for different combinations of bose and fermi 
    multipole ($L$) operators with $L=0,1,2,3$. Transition densities and 
    dynamical structure factors are calculated  for collective excitations. 
    Comparison with the sum rule prediction for the collective frequency 
    is given. Time dependent behavior of the system 
    after an external impulse is studied. 
    \end{abstract}


\section{introduction}

Collective excitation is one of the most prominent phenomena in quatum 
many-body systems such as liquid helium, electron gas,  nucleus, etc. In 
the recently developed Bose-Einstein condensates of trapped atomic 
gases\cite{bec,Dalfovo}, collective oscillations were the first of the dynamical 
phenomena discovered\cite{osc}. Collective oscillations are characterized 
by various quantum numbers related to, e.g., the shape of oscillation, or 
internal degrees of freedom of the constituents such as spin, isospin, etc. 
The oscillation frequency, the damping width, etc. depend on the interparticle 
interaction of the constituents, and thus provide a clue to unravel the 
dynamical correlation of the many-body system. 

Study of the properties of trapped neutral atoms has  been 
extended to fermi systems\cite{Demarco} where the occurrence of the
Fermi degeneracy was observed, and also to the mixture of bose and 
fermi particles. The latter system with condensed bosons and degenerate 
fermions  was recently realized experimentally \cite{Rice,EPS}. This system is 
one of the typical example in which particles obeying different 
statistics are intermingled. Theoretical studies of the bose-fermi mixed 
system of cold atomic gases have been done for  static 
properties\cite{Molmer,Amoruso,MOSY,Bijlsma,Vichi,Vichi1}, 
for the phase diagram and phase separation\cite{Nygaard,Yi,Viverit}, for the 
stability of the system\cite{MSY1,Roth} and for  collective 
excitations\cite{MSY,Minguzzi,Capuzzi}. 

In Ref.~\cite{MSY} sum rule approach has been applied for collective 
excitations in the bose-fermi mixed system. Average excitation energies 
for the states with multipoles $L=0,1,2$ were calculated for both 
in-phase and out-of-phase modes of the bose and the fermi particles, 
and the dependence on the bose-fermi interaction strength was studied. 
The sum rule approach is a powerful technique for collective states in 
quantum many-body systems, and has been successfully applied to the 
excitation of bose condensed systems, see Refs.~\cite{Dalfovo,sum}. 
It does not give, however, direct information on the eigenstates of the 
system, but rather an average behavior of the strength distribution for 
the adopted multipole excitation operator. For a more detailed 
investigation on the dynamical properties of the system, one would need 
a study of individual eigenstates. 

In the present paper we extend the previous study\cite{MSY} of collective 
excitations in the bose-fermi mixed system. We  calculate full 
particle-hole type excitations by 
diagonalization of an RPA type matrix. The motive of this 
calculation is threefold: first of all, the calculation allows us to 
study the excitation spectrum and the strength distribution 
in contrast to the sum rule method which 
focuses on the strength weighted average of the prescribed multipole 
operators. We study, for instance,  the degree of collectivity of the 
excitations depending on the strength of boson-fermion interaction, 
and estimate the damping of the collective excitation albeit within the 
space of particle-hole excitations. Information on the wave function 
allows us to calculate observables such as the dynamical structure 
factor. The latter for bose-condensed systems is now becoming available 
experimentally by two-photon spectroscopy\cite{twophoton} and is 
being studied theoretically\cite{response}. 
Secondly, we compare the results with those obtained from the sum rule 
approach. This provides a check on the approximation 
adopted in the calculation such as the model space truncation. The 
comparison also allows one to examine the structure of the low- and 
high-lying collective modes which was speculated in \cite{MSY} through 
the mixing angle $\theta$ of multipole operators. 
Third, we can predict the 
time-dependent behavior of the system for a given external perturbation. 
This process is actually the one that 
has been employed in the previous experimental study of collective 
excitations in BEC. An RPA study of the bose-fermi mixed system has 
recently been done in \cite{Capuzzi}, where the response  for 
an external multipole field is formulated in the form of 
an integral equation. In the present 
calculation we approach the problem by diagonalizing the RPA matrix, 
by incorporating the discrete nature of the excitation in an 
isolated trapped system, and investigate, 
in particular, the properties of the strength distribution for 
a combination of bose and fermi operators and for various values of the 
bose-fermi interaction strengths.  


The content of the paper is as follows: in the next section we derive 
the RPA equation for the bose-fermi mixed system using the equation of 
motion for  particle-hole type excitation operators. The single 
particle (hole) states are obtained in the mean-field calculation, 
i.e., by solving the coupled Gross-Pitaevskii and Thomas-Fermi equations. 
In Section \ref{sec:calculation} we first  briefly describe the parameters  and the numerical 
procedure employed in the present calculation. 
We then turn to the detailed studies of the results obtained, 
including ground state density, single particle states, and strength 
distribution for each multipole. Comparison with the sum rule calculation 
is given. Transition densities and dynamical structure factors for 
some of the collective excitations are presented. We finally 
consider the time development of the system after an 
external multipole impulse on the system. Last section 
is devoted to summary and conclusion. 

\section{formulation}
We consider a dilute spin-polarized bose-fermi mixed system trapped 
in a spherically symmetric harmonic oscillator potential at $T=0$. 
The system is described by the Hamiltonian
\begin{equation}
      \hat{H} =\hat{H}_0+\hat{V}_b+\hat{V}_{bf}
      \end{equation}
with
\begin{eqnarray}
      \hat{H}_0&=& \int \!\! d^3 r \hat\psi^\dagger(\vec r)
         \left[- \frac{\hbar^2}{2m}\vec{\nabla}^2 
         + \frac{1}{2}m\omega_0^2 r^2\right] \hat\psi(\vec r) 
         + \int \!\! d^3 r \hat\phi^\dagger(\vec r)
         \left[- \frac{\hbar^2}{2m}\vec{\nabla}^2 
         + \frac{1}{2}m\omega_0^2 r^2\right]\hat\phi(\vec r), \nonumber \\
      \hat{V}_b&=& \frac{1}{2} g \int \!\!\! \int \!\! d^3 r d^3 r' 
         \hat\phi^\dagger(\vec r) \hat\phi^\dagger(\vec r')
         \delta^{(3)}(\vec r - \vec r')
         \hat\phi(\vec r') \hat\phi(\vec r), \nonumber \\
      \hat{V}_{bf}&=& h \int \!\!\! \int \!\! d^3 r d^3 r' 
         \hat\phi^\dagger(\vec r) \hat\psi^\dagger(\vec r')
         \delta^{(3)}(\vec r - \vec r')
         \hat\psi(\vec r') \hat\phi(\vec r) \mbox{,}
      \end{eqnarray}
where $\hat\phi$ and $\hat\psi$ are the boson and the fermion field operators.
We take, for simplicity,  the mass $m$ and the trapping 
frequency $\omega_0$ to be the same for the boson and the fermion. 
The boson-boson and the boson-fermion interaction strengths for 
the pseudopotentials, $g$ and $h$, are 
related to the $s$-wave scattering lengths $a_{bb}$ and $a_{bf}$ 
through $g=4\pi\hbar^2a_{bb}/m$, $h=4\pi\hbar^2a_{bf}/m$, while the 
fermion-fermion interaction is omitted as we consider a polarized dilute 
system at low temperature.
\vspace{2mm}

According to the Landau picture of quantum liquid, low lying excited 
states of the system may be described by quasiparticle excitations which 
have the structure of particle-hole 
(p-h) type excitations from the ground state. Small amplitude collective 
oscillations of the system, corresponding to the zero sound, are given by 
the coherent superposition of these p-h excitations and are well described 
by RPA for many fermion systems. The corresponding excitations in the bose 
condensed system have been formulated in the Bogoliubov-de Gennes type 
equation. The latter is essentially the same as the one obtained in RPA,
and we consider below the two types of excitations on the same footing. 

In order to formulate RPA for the bose-fermi mixed system, we first determine 
the ground state in the mean field approximation. The obtained mean field 
provides us with single particle energies and wave functions. Let us 
expand the field operators in terms of a complete set of single particle 
wave functions as
\begin{equation}
  \hat\phi(\vec r)=\sum_k \phi_k(\vec r)b_k,
  \qquad
  \hat\psi(\vec r)=\sum_\alpha \psi_\alpha(\vec r)a_\alpha,
  \end{equation}
where  $b_k$ and $a_\alpha$  are the boson and the fermion annihilation 
operators in the single particle states specified by the wave functions 
$\phi_k$ and $\psi_\alpha$. They satisfy the standard commutation or 
anticommutation relations. The single particle states with quantum numbers 
$k$ or $\alpha$ are determined by minimizing the energy expectation value. 
The trial wave function of the system is given by the product of $N_b$ bosons 
in a single 
state $\phi_{k=0}(\vec r)$ and the Slater determinant of $N_f$ different 
fermionic states $\psi_\alpha(\vec r)$. From the stationary condition of 
energy for the variation of these wave functions with a given number of 
bosons $N_b$ and fermions $N_f$, we obtain the set of coupled equations, 
i.e., the Gross-Pitaevskii equation for the boson wave function
\begin{equation}
      \left[ -\frac{\hbar^2}{2m}\nabla^2 + \frac{1}{2}m\omega_0^2 r^2
      +gN_b|\phi_0(\vec r)|^2 +h\rho_f(\vec r)
      \right]\phi_0(\vec r) = \mu_b \phi_0(\vec r)
      \label{GPeq}
\end{equation}
and the mean-field equation for the occupied fermion states
\begin{equation}
      \left[ -\frac{\hbar^2}{2m}\nabla^2 + \frac{1}{2}m\omega_0^2 r^2
      +hN_b|\phi_0(\vec r)|^2 \right]\psi_\alpha(\vec r) = \epsilon^f_\alpha
      \psi_\alpha(\vec r),
      \label{HFeq}
      \end{equation}
where 
\begin{equation}
   \rho_f(\vec r)=\sum_\alpha^{\hbox{occupied}}|\psi_\alpha(\vec r)|^2
   \label{rhof}
   \end{equation}
is the ground state density of  fermions, $\mu_b$ is the boson 
chemical potential and $\epsilon^f_\alpha$ are the single particle 
energies for the fermion. Set of equations (\ref{GPeq}) and (\ref{HFeq}) 
together with the fermion number condition
\begin{equation}
    \int\!d^3r\,\rho_f({\vec r}) =N_f
    \end{equation}
constitute a closed set of equations. The single particle wave functions are 
normalized to unity. 
In Eq.~(\ref{GPeq}) we approximate the factor $N_b-1$ by $N_b$. 
Once the single particle wave functions 
for the occupied states are obtained from Eqs.~(\ref{GPeq}) and 
(\ref{HFeq}), the wave functions of the unoccupied states are 
calculated from a similar set of equations, i.e., 
\begin{equation}
      \left[ -\frac{\hbar^2}{2m}\nabla^2 + \frac{1}{2}m\omega_0^2 r^2
      +gN_b|\phi_0(\vec r)|^2 +h\rho_f(\vec r)
      \right]\phi_k(\vec r) = \epsilon_k^b \phi_k(\vec r)
      \label{GPunocc}
\end{equation}
for bosons with $k\ne 0$, and the same equation (\ref{HFeq}) for fermions 
but for 
unoccupied states. Orthogonality of these wave functions to the occupied 
ones are automatically satisfied \cite{Ring}. 

For a large number of particles with a smooth mean field potential, the 
Thomas-Fermi calculation provides a good approximation. Assuming this 
to be the case, we determine the {\it fermionic} ground state density 
$\rho_f$ by using 
\begin{equation}
     \frac{\hbar^2}{2m}(6\pi^2\rho_f(\vec r))^{2/3}+\frac{1}{2}m
     \omega_0^2 r^2+hN_b|\phi_0(\vec r)|^2=\epsilon_F
     \label{TFapp}
     \end{equation}
together with the equation (\ref{GPeq}) instead of fully solving the 
coupled set of equations (\ref{GPeq}) and (\ref{HFeq}). The 
Fermi energy $\epsilon_F$ is determined by integrating the density 
$\rho_f(\vec r)$ so as to give the fermion number.  Numerical 
consistency of this procedure will be shown in the next section. 
With $\rho_f$ and $\phi_0$ so determined, the wave functions 
$\phi_k\,(k\ne 0)$ and $\psi_\alpha$ are obtained as described above. 

We assume below that the mean field  is spherically symmetric, 
and the single-particle quantum numbers $\alpha$ and $k$ involve 
standard radial and angular momentum quantum numbers $(n,\ell,m)$. 
(Note that 
the spin quantum number is frozen.) We consider the case in which 
the fermionic ground state (Slater determinant) is $m$-closed and thus 
is consistent with the spherically symmetric mean field.  
We denote by $p$ ($h$) the fermion single particle states unoccupied (occupied) 
in the Slater determinant, i.e., those above (below) the Fermi 
energy $\epsilon_F$. For the bosons at $T=0$ the occupied state is the 
lowest single-particle state $k=0$.

\vspace{2mm}

We now define the creation and annihilation multipole operators of the 
excited state $|\nu L M \rangle$ with angular momentum quantum numbers $L,M$ 
and an additional quantum number $\nu$:
\begin{eqnarray}
   Q^\dagger_{\nu LM}&=&F^\dagger_{\nu LM}+B^\dagger_{\nu LM}, 
   \label{eqn.10}\\
   F^\dagger_{\nu LM}
   &=&\sum_{ph}\sum_{m_p m_h}\left[
   X^\nu_{phL}\langle l_p m_p l_h -m_h \,|\, L\,M\,\rangle(-1)^{l_h-m_h}
   a^\dagger_{p l_p m_p}a_{h l_h m_h} \right. \nonumber \\
   &&\left.-Y^\nu_{phL}\langle l_h m_h l_p -m_p \,|\, L\,M\,\rangle(-1)^{l_p-m_p}
   a^\dagger_{h l_h m_h}a_{p l_p m_p}
   \right],\\
   B^\dagger_{\nu LM}
   &=&\frac{1}{\sqrt{N_b}}\sum_{k \neq 0}\left[
   U^\nu_{kL}b^\dagger_{kLM}b_0-V^\nu_{kL}(-1)^{L+M}b^\dagger_0b_{kL-M}
   \right],
   \label{eqn.12}
\end{eqnarray}
where $X,Y,U,V$ are the amplitudes to be determined in RPA, 
and $\langle l_p m_p l_h -m_h \,|\, L\,M\,\rangle$ is the 
Clebsh-Gordan coefficient. 

In RPA the operator $Q$ is determined so as to satisfy the equation of motion
\begin{equation}
   [H,Q^\dagger_{\nu LM}]\simeq \hbar \Omega_\nu Q^\dagger_{\nu LM}, 
   \label{eqn.14}
\end{equation}
where the terms in the l.~h.~s. which involve  different combination of 
operators $a^\dagger, a, b^\dagger, b$ from that in  $Q^\dagger$ are 
neglected by the assumption 
of RPA. The state 
\begin{equation}
   |\nu LM\rangle=Q^\dagger_{\nu LM}|0\rangle\qquad
   \hbox{with}\qquad   Q_{\nu LM}|0\rangle=0
\end{equation}
is then an approximate eigenstate of the Hamiltonian together with 
the correlated 
ground state  given by $|0\rangle$. The amplitudes satisfy the 
orthonormality condition
\begin{equation}
   \sum_{ph}(X^{\nu *}_{phL}X^{\nu'}_{phL}-Y^{\nu *}_{phL}
   Y^{\nu'}_{phL})
   +\sum_{k\ne0}(U^{\nu *}_{kL}U^{\nu'}_{kL}-V^{\nu *}_{kL}V^{\nu'}_{kL})
   =\delta_{\nu\nu'}.
\end{equation}

Substituting Eq.~(\ref{eqn.10}) into Eq.~(\ref{eqn.14}) 
we obtain the eigenvalue equation in matrix form:
\begin{eqnarray}
   &&\left(\begin{array}{rrrr}
   A_{XX} & A_{XU} & B_{XY} & B_{XV} \\
   A_{UX} & A_{UU} & B_{UY} & B_{UV} \\
   -B_{XY} & -B_{XV} & -A_{XX} & -A_{XU} \\
   -B_{UY} & -B_{UV} & -A_{UX} & -A_{UU}
   \end{array}\right)
   \left(\begin{array}{c}
   X \\ U \\ Y \\ V
   \end{array}\right)
   =\hbar \Omega_\nu
   \left(\begin{array}{c}
   X \\ U \\ Y \\ V
   \end{array}\right),
\end{eqnarray}
where submatrices $A$ and $B$ are given by
\begin{eqnarray}
   &&(A_{XX})_{ph,p'h'}
   =[\epsilon^f_{p l_p}-\epsilon^f_{h l_h}]
   \delta_{p,p'}\delta_{h,h'}\delta_{l_p,l_p'}\delta_{l_h,l_h'}, \nonumber \\
   &&(A_{XU})_{ph,k}=
   \frac{h\sqrt{N_b}}{4\pi}\frac{\hat{l_p}\hat{l_h}}{\hat{L}}(-1)^{l_h}
   \langle l_p 0 l_h 0 | L 0 \rangle
   \int\!\! dr r^2 R^f_{pl_p} R^f_{hl_h}R^b_{kL}R^b_{00}, \nonumber \\
   &&(A_{UX})_{k,ph}=(A_{XU})_{ph,k},  \nonumber \\
   &&(A_{UU})_{k,k'}=
   [\epsilon^b_{kL}-\mu_b]\delta_{kk'}
   +\frac{gN_b}{4\pi}\int\!\! dr r^2 R^b_{kL} R^b_{k'L}R^b_{00}R^b_{00}, \nonumber \\
   &&(B_{XY})_{ph,p'h'}=0,  \label{submatrix} \\
   &&(B_{XV})_{ph,k}=
   \frac{h\sqrt{N_b}}{4\pi}\frac{\hat{l_p}\hat{l_h}}{\hat{L}}(-1)^{l_p}
   \langle l_p 0 l_h 0 | L 0 \rangle
   \int\!\! dr r^2 R^f_{pl_p} R^f_{hl_h}R^b_{kL}R^b_{00}, \nonumber \\
   &&(B_{UY})_{k,ph}=(B_{XV})_{ph,k},\nonumber\\
   &&(B_{UV})_{k,k'}=
   \frac{gN_b}{4\pi}(-1)^L\int\!\! dr r^2 R^b_{kL} R^b_{k'L}R^b_{00}R^b_{00} \nonumber
   \end{eqnarray}
with $ \hat l = \sqrt{2l+1}$. 
In Eq.~(\ref{submatrix}) $R^b_{nl}(r)$ and $R^f_{nl}(r)$ are the radial parts of 
the single particle wave functions $\phi$ and $\psi$,
which are defined by $\phi_{nlm}(\vec r)=R^b_{nl}(r)Y_{lm}(\theta,\varphi)$ 
and $\psi_{nlm}(\vec r)=R^f_{nl}(r)Y_{lm}(\theta,\varphi)$ 
where $Y_{lm}(\theta,\varphi)$ are the spherical harmonics.
We omitted  $O(N_b^{-1})$ terms in this calculation. Actually we obtain the 
same set of equations if we replace $b^\dagger_0,b_0$ in Eq.~(\ref{eqn.12}) 
with $\sqrt{N_b}$, which is equivalent to the Bogoliubov-de Gennes 
equation for the bosonic system. 

\vspace{2mm}
The response of the system to an external field is represented by the 
transition matrix elements of the  relevant operators which connect the ground 
state and an excited state. The external probe for a collective oscillation 
is assumed to be long-ranged and is given by the standard multipole 
operators, i.e., 
\begin{equation}
    F_L(\vec r)=   \left\{
    \begin{array}{ll}
    \displaystyle  \frac{1}{\sqrt{4\pi}}r^2 & (L = 0) \\
    \displaystyle  r^L Y_{L0}(\theta) & (L\neq 0)
    \end{array} \right. .
    \label{Fmul}
\end{equation}
As the system is composed of two kinds of particles, we define operators
\begin{equation}
     F_L^b=\int\!d^3r \hat\phi^\dagger(\vec r)F_L(\vec r)\hat\phi(\vec r),\qquad
        F_L^f=\int\!d^3r \hat\psi^\dagger(\vec r)F_L(\vec r)\hat\psi(\vec r) 
        \end{equation}
and their combination
\begin{equation}
    F^\pm_L=F^f_L \pm F^b_L ,
\end{equation}
where $F_L^\tau$ with $\tau=f,b,+,-$  are resepectively called 
`fermionic', `bosonic', `in-phase' and `out-of-phase' type operators. 
The transition amplitudes for these operators are calculated in RPA as:
\begin{eqnarray}
   &&\langle 0 | F^\pm_L | \nu L0 \rangle 
   =\langle 0 | [F^\pm_L,Q^\dagger_{\nu L0}] | 0 \rangle 
   =\langle 0 | [F^f_L,F^\dagger_{\nu L0}] | 0 \rangle 
   \pm \langle 0 | [F^b_L,B^\dagger_{\nu L0}] | 0 \rangle, \\
   &&\langle 0 | [F^f_L,F^\dagger_{\nu L0}] | 0 \rangle 
   =\sum_{ph} \frac{1}{\sqrt{4\pi}} \frac{\hat{l_p}\hat{l_h}}{\hat{L}}
   \langle l_p 0 l_h 0 | L 0 \rangle
   [(-1)^{l_h}X^\nu_{phL}+(-1)^{l_p}Y^\nu_{phL}]
   \int\!\! dr r^2 R^f_{h l_h}r^\lambda R^f_{p l_p},\\
   &&\langle 0 | [F^b_L,B^\dagger_{\nu L0}] | 0 \rangle 
   =\sum_{k} \frac{\sqrt{N_b}}{\sqrt{4\pi}}
   [U^\nu_{phL}+(-1)^{L}V^\nu_{phL}]
   \int\!\! dr r^2 R^b_{k L}r^\lambda R^b_{00},
\end{eqnarray}
where $\lambda=2$ for $L=0$ and $\lambda=L$ for $L\neq 0$.
The strength distribution for the multipole operator $F_L^\tau$  is 
then given by
\begin{equation}
      \sum_\nu \delta(\Omega-\Omega_\nu)|\langle 0|F_L^\tau|\nu L0\rangle|^2.
      \qquad (\tau=f,b,+,-)
      \end{equation}
They measure the collectivity of the excited states with 
respect to the multipole operator which characterizes the shape 
(or volume for $L=0$) oscillation of the system. 

\vspace{2mm}

More detailed information on the structure of the collective excitation 
may well by represented by the transition densities\cite{Serr}
\begin{eqnarray}
    &&\langle 0 | \hat{\rho}^\tau(\vec r) | \nu LM\rangle
    =\delta\rho^\tau_{\nu L}(r) Y_{LM}(\theta,\varphi) 
    \quad (\tau=f,b),\\
    &&\delta\rho^f_{\nu L}(r)
    =\sum_{ph} \frac{1}{\sqrt{4\pi}} \frac{\hat{l_p}\hat{l_h}}{\hat{L}}
    \langle l_p 0 l_h 0 | L 0 \rangle
    [(-1)^{l_h}X^\nu_{phL}+(-1)^{l_p}Y^\nu_{phL}] R^f_{h l_h}R^f_{p l_p}, \\
    &&\delta\rho^b_{\nu L}(r)
    =\sum_{k} \frac{\sqrt{N_b}}{\sqrt{4\pi}}
   [U^\nu_{phL}+(-1)^{L}V^\nu_{phL}]
   R^b_{k L}R^b_{00},
\end{eqnarray}
where $\hat{\rho}^f(\vec r)$,$\hat{\rho}^b(\vec r)$ 
are the fermion and boson density operators:
\begin{equation}
    \hat{\rho}^f(\vec r)=\hat\psi^\dagger(\vec r)\hat\psi(\vec r),\qquad
    \hat{\rho}^b(\vec r)=\hat\phi^\dagger(\vec r)\hat\phi(\vec r).
    \end{equation}
Recent development of the two-photon Bragg spectroscopy\cite{twophoton} 
allows us to obtain dynamical structure factors (response function) 
for the excitations in trapped atomic gases. They are related to 
the transition densities as
\begin{eqnarray}
   &&S_L^\tau(q,\Omega)
     =\sum_\nu|{\cal F}_{\nu L}^\tau(q)|^2\delta(\Omega-\Omega_\nu), \\
   &&{\cal F}_{\nu L}^\tau(q)=\int\!dr\,r^2j_L(qr/\hbar)\,\delta\rho_{\nu L}^\tau(r),
   \end{eqnarray}
where $q$ denotes momentum transferred to the system and 
$j_L(qr/\hbar)$ are spherical Bessel functions of order $L$. 
The long wavelength limit of the dynamical structure factor is 
proportional to the strength distribution for the multipole operators 
defined above.

\section{Calculation}
\label{sec:calculation}
\subsection{Numerical procedure}

We consider a boson-fermion mixture of potassium isotopes, i.e., 
$^{41}$K (boson) and $^{40}$K (fermion).
We take the same value $m=0.649\times10^{-25}$kg 
for the boson and the fermion masses. 
The boson-boson interaction strength $g$ is obtained from the scattering length 
$a_{bb}=15.13$nm for $^{41}$K-$^{41}$K\cite{abb}, 
while the boson-fermion interaction strength $h$ is varied. 
(A negative value for the scattering length $a_{bf}$ for $^{40}$K-$^{41}$K 
has been suggested in \cite{abb}, although not well established yet.) 
Both particles are assumed to be trapped in the spherical 
oscillator potential with the same trap frequency $\omega_0=100$Hz. The 
oscillator length $\xi=\sqrt{\hbar/m\omega_0}$ is 4.03$\mu$m for the 
adopted value of $\omega_0$.

Number of bosons is fixed at $N_b=$1000, while that of
fermions is calculated in the Thomas-Fermi approximation 
for a fixed Fermi energy $\epsilon_F=(6 \times 1140)^{1/3}\hbar\omega_0$. 
The latter is determined  so that the last oscillator shell with a number 
of quanta $N=2n+\ell$ is given by $N_{\rm Fermi}=17$ at $h/g=0$. This gives the 
number of fermions $N_f=1140$ at $h/g=0$. $N_f$ is dependent on 
the boson-fermion interaction strength, and it is assumed that all 
the subshells given by the quantum numbers $(n,\ell)$ are $m$-closed, 
which gives, for instance, $N_f=1050$ at $h/g=8$ and  $N_f=1227$ 
at $h/g=-6$ for the given value of $\epsilon_F$.

The single-particle wave functions were obtained from the coupled 
Gross-Pitaevskii and Thomas-Fermi equations by expanding the 
wave functions in the harmonic oscillator basis for the given oscillator 
constant $\xi$. We included  associated Laguerre functions $L^\alpha_n$ 
up to $n=30$ for fermions and $n=15$ for bosons. 

The excitation energies and the wave functions for each multipole $L$
are obtained by diagonalization of the RPA matrix. To construct the 
particle-hole basis we included single particle states for bosons up to 
$n=7$ for $\ell=0$ and $n=6$ for $\ell=1,2,3$, so that seven p-h configurations 
are included for each $L=0,1,2,3$. For fermions, four (particle) states 
above the Fermi level $\epsilon_F$ and up to four (hole) states 
below $\epsilon_F$ have been included for each $\ell$.
The number of fermi p-h configurations is 240 for monopole,
464 for dipole, 672 for quadrupole and 1124 for octupole.
The dimension of the RPA matrix is twice  the sum of bose and fermi 
configurations for each multipole. The number of bose configurations 
is the same for the four multipoles considered in the present 
calculation.

\subsection{Static properties}

In the ground state the fermion density distribution is 
much broader than that for the boson due to Fermi pressure \cite{MOSY,Roth}, 
although  at large negative value of $h/g$  the bose-fermi attraction 
tends to produce a larger overlap of the two kinds of particles 
as shown in Ref.~\cite{Roth}. 
As the bose-fermi interaction 
becomes repulsive the fermions are squeezed out from the 
central part. This in turn causes a smaller overlap of 
bosons and fermions, and thus a relatively small 
net effect of the interaction on the binding energy. 
For the present choice of parameters the fermions are 
distributed outside of the boson `core' around $h/g\simeq 7$, 
forming a `shell' like structure \cite{Molmer,Amoruso,Nygaard,Roth}. 
Note that this behavior would be changed if one 
adopts different values for $N_f/N_b$ and $g$, which 
may be represented by a single parameter $\alpha$ introduced in 
Ref.~\cite{MOSY}. 

The condition for the validity of the Thomas-Fermi approximation 
used to obtain the above fermion density distribution may be 
expressed in terms of the local de Brogile wave length 
$\lambda(r)=\hbar/p(r)$, 
where $p(r)=\sqrt{2m(\epsilon_F-V_{\rm eff})}$ with 
fermion mean field potential 
$V_{\rm eff}(r)=\frac{1}{2}m\omega_0^2r^2+h\rho_b(r)$.
With this quantity, the condition becomes \cite{Landau}
\begin{equation}
f(r)=\left| \frac{d\lambda(r)}{dr} \right| \ll 1.
\end{equation}
At $h/g=8$ where the fermionic potential may have a most 
pronounced structure, the value of $f(r)$ is $\sim 10^{-2}$ 
except around turning points.
At the turning points, however, the fermion density produces only
a negligible influence on the mean field potential.
The validity of the present Thomas-Fermi calculation 
may also be checked by comparing the fermion density distribution 
$\rho_f(r)$ obtained from Eq.~(\ref{TFapp}) with 
the one calculation from the single particle wave functions 
according to Eq.~(\ref{HFeq}).
The comparison of the two distribution in the range $h/g=-6\sim8$ 
shows that they agree within the order of $10^{-2}$ 
over the entire radial range,
suggesting the consistency of the present calculation.

In Fig.~\ref{fig:FSPE} we show the fermion single-particle energies measured 
from Fermi energy, $\epsilon_\alpha-\epsilon_F$, 
against orbital angular momentum $\ell$ for three values of 
the interaction parameter, $h/g=5.0,1.0, -3.0$. The Fermi energy 
$\epsilon_F$ is denoted by the horizontal line at zero energy. Note that 
$\epsilon_\alpha$ are given by those of the harmonic oscillator at $h/g=0$, 
$\epsilon_{n,\ell}=(2n+\ell+3/2)\hbar\omega_0$, as 
we have no direct interaction among fermions. One should 
note also that the yrast state, i.e., the lowest state for each 
angular momentum, has no radial node, and the yrare one, the 
second lowest state, has one radial node, etc. The nodal 
structure of the particle and hole states around the Fermi 
surface is responsible for the multipole strength distribution of low energy 
particle-hole excitations. 
The figure shows that the states with low orbital angular momentum 
are much influenced by the bose-fermi interaction, while those 
with high angular momentum are almost insensitive to the 
values of $h/g$. This is because the additional fermion 
potential due to the bose-fermi interaction vanishes outside 
the boson density distribution. For instance, since the yrast 
single particle wave functions with angular momentum $\ell$ 
are peaked around $\sqrt{\ell}\xi$, those fermion states with 
$\ell> (R_B/\xi)^2$, $R_B$ being a typical edge radius of the boson 
distribution,  would not be much influenced by the 
bose-fermi interaction. (In the present case $R_B\simeq 3\xi$, 
and the above relation gives $\ell>9$.)

\subsection{Energy weighted moments and comparison with the sum rule calculation}

RPA calculation provides  approximate eigenvalues and wave functions for 
all the individual eigenstates and is useful to study details of the 
dynamics of the system. If one is interested in the gross behavior of 
the strength distribution or an approximate frequency of the oscillation, 
one may rather consider the $p$-th energy weighted moments of the 
strength distribution,
\begin{equation}
   m_p(L,\tau)=\sum_\nu (\hbar \Omega_\nu)^p |\langle 0|F_L^\tau| \nu \rangle|^2,
   \label{moment}
   \end{equation}
where $\hbar \Omega_\nu$ is the excitation energy of the state $|\nu\rangle$. 
These moments can be expressed as ground-state expectation values 
of  multiple commutators of $F_L^\tau$ with the 
Hamiltonian\cite{Bohigas,Lipparini}. It is known that this relation, 
sum rule, is conserved in RPA for some of the moments, i.e.,  the 
sum in the r.h.s. of Eq.~(\ref{moment}) obtained from RPA eigenstates 
coincides with the expectation value of the  commutator in 
the HF ground state\cite{Thouless,Bohigas}. 
Thus a comparison of the two would provide a criterion on the consistency 
of the RPA calculation. In particular, the first moment (energy weighted 
sum, EWS) $m_1$ is calculated from the double commutator of $F_L^\tau$ 
and the Hamiltonian as 
\begin{eqnarray}
  m_1(L,\tau)=\frac{1}{2}\langle [F_L^\tau,[\hat{H},F_L^\tau]]\rangle_0 ,
      \label{sumrule}
      \end{eqnarray}
where $\langle\;\;\rangle_0$ denotes a ground state expectation 
value\cite{Bohigas}. 
We have checked that the  sum rule is satisfied within $1$\% for all 
the multipole modes. 

One can estimate the average frequency of the collective oscillation 
based on sum rules. There are several ways to define the average 
frequency depending on which part of the strength distribution 
is emphasized. In Ref.~\cite{MSY}  the ratio of the third and 
the first moments 
\begin{equation}
   \bar{\omega} = \sqrt{\frac{m_3}{m_1}}
  \label{average}
  \end{equation}
was studied based on sum rules. This definition is advantageous as 
it can be used to test the validity of the RPA calculation as 
mentioned earlier. 
In  later discussions we consider also the average frequency
calculated by
\begin{equation}
    \omega_{av}=\frac{m_1}{m_0}
    \end{equation}
which has more weight in the low-frequency strength compared with 
$\bar\omega$. The two frequencies should coincide when a single 
collective state exhausts the strength. We discuss later also the 
width of the strength distribution defined by 
\begin{equation}
    \sigma=\left[\frac{m_2}{m_0}-\left(\frac{m_1}{m_0}\right)^2\right]^{1/2}.
    \end{equation}

In Fig.~\ref{fig:AFCE} we show the average frequency Eq.~(\ref{average}) for $L=0,1,2$ 
and $\tau=+,-$ given by the sum rule (\ref{sumrule}) and the one 
directly calculated in RPA. We find a good agreement of the two 
calculations in a wide range of the bose-fermi interaction 
parameter $h$ for a fixed $g$. There is a discrepancy at very 
large values of $|h|$, especially in the  strongly attractive case. 
In the latter case, an induced instability of the ground state 
due to the bose-fermi attraction may be close\cite{MSY1}, which suggests 
that a larger configuration space 
may be required to satisfy the sum rule. In fact, by increasing 
the number of particle-hole states, we obtained a 
better agreement.

\subsection{Distribution of multipole strengths}

Now we show the distributions of the multipole strengths. 
Figures \ref{fig:MSD-FB},\ref{fig:DSD-IO},\ref{fig:QSD-FB},\ref{fig:OSD-FB} 
respectively show the strength distributions for $L=$0,1,2 and 3,
for either the fermionic/bosonic  or the 
in-phase/out-of-phase operators.

\subsubsection{monopole}

In Fig.~\ref{fig:MSD-FB} we show the monopole strength distribution,  
$|\langle{0}|F_0^\tau |{\nu}\rangle|^2$ against $\Omega_\nu$, for 
$\tau=f,b$ and for three values of $h/g$. 
Here one expects a volume oscillation of boson and/or fermion 
densities. For $h=0$ and for a large number of particles, the bosonic 
monopole oscillation will be located around $\sqrt{5}\omega_0$ 
as given by the collisionless hydrodynamics\cite{Dalfovo,sum}, 
while the fermionic one is concentrated at $2\omega_0$. 
The latter property is due to the almost degenerate values of 
the relevant particle-hole energies contributing to the 
monopole oscillation. 
The present calculation show that this situation persists 
even for large values of $h/g$, although the bosonic 
frequency is slightly shifted. The fermionic strength is 
distributed over a few states around $2\omega_0$, showing that the induced 
fermi-fermi interaction via bosons is not strong enough to 
make a coherent superposition of the fermion particle-hole 
states. 
The total monopole strength for 
fermion is by an order of magnitude larger than that for  boson, 
even though the number of fermions involved in the excitation 
is smaller because of the Pauli principle, e.g. 
$m_1=2590\hbar\omega_0\xi^4$ for fermion and 
$m_1=419\hbar\omega_0\xi^4$ for boson at $h/g=1$. 
This is because of the broader density distribution for fermions.

The strength distribution may suggest that the fermions and bosons are 
moving independently even for rather large values of $h/g$. This is 
not necessarily the case, however,  as one may see in Fig.~\ref{fig:DSF-M}, where 
we plot the dynamical structure factor for the three cases, i.e., the 
states at $\Omega/\omega_0=2.03$ for $h/g=5.0$, at
$2.01$ for $h/g=1.0$ and at $1.98$ for $h/g=$-3.0, which carry the 
largest strengths for the in-phase monopole operator. Although the 
fermionic  strength is far larger than the bosonic one for 
this state, the mixture of the bosonic component is not small and is 
peaked at larger values of the momentum transfer $q$ as shown by the 
dashed lines. The latter 
shows that the bosons oscillate in the inner region although not 
quite recognizable as far as one studies only  the $F_0$ distribution.

In Refs.~\cite{MSY1,MSY} we suggested that 
an instablity towards collapse may occur at large negative values of $h/g$ 
and may be signaled by the lowering of $\bar\omega$. In the present calculation 
with smaller number of particles this is not apparent in Fig.~\ref{fig:AFCE}. In 
Fig.~\ref{fig:SDIM} we show the strength distribution at $h/g=-6.65$. 
This value is close to the critical value 
of instability around $-6.7$ estimated from the 
M{\o}lmer's condition~\cite{Molmer,MSY1} and 
around $-6.8$ from the condition by Roth and Feldmeier~\cite{Roth}.
We find a lowering 
of a single state which carries 13.9\% of the EWS. The average frequency for 
this case is $\bar\omega=1.92\omega_0$ appreciably lower 
than the value $\sim 2\omega_0$ for $h/g=-5\sim 5$. At more 
negative values of $h/g$ we could not find a stable ground state. 

Let us study the character of the low-lying excited states by 
considering the response to the probe
\begin{equation}
   F(\theta)=F^+\cos\theta+F^-\sin\theta
   \end{equation}
parametrized by $\theta$. In Ref.~\cite{MSY} the angle $\theta$ was 
determined by minimizing the average frequency $\bar\omega$ for $F(\theta)$ so 
as to find the character of the probe which favors the low-lying 
states. Once the value of the parameter, $\theta_{\rm min}$ is 
determined for a given $h/g$, one may consider an operator 
`perpendicular' to $F(\theta_{\rm min}$), i.e., 
$F_\perp=F^+\sin\theta_{\rm min}-F^-\cos\theta_{\rm min}$, which 
may favor the high-lying state. Alternatively, one may maximize 
the EWS within the model space to determine the value $\theta_{\rm max}$ 
of the operator to characterize the high-lying states. Figure \ref{fig:SD-theta} shows 
the strength distribution for these three types of  operators 
at $h/g=5.0$. The value $\theta_{\rm min}=-0.12\pi$ suggests that, at this 
highly repulsive value of the interaction, a slightly in-phase type oscillation 
is favored in order to avoid an overlap of bosons and fermions, see the 
density distribution in Ref~\cite{Molmer,Amoruso,Nygaard,Roth}. 
We note that the distribution for the 
`perpendicular' operator $F_\perp$ is concentrated and is similar to the one 
for $F(\theta_{\rm max})$. The latter, however, is an almost bosonic 
type operator and suggest that the strength distribution alone is 
not sufficient to characterize the structure of high-lying states. 
A similar study has been made for a very attractive case $h/g=-6.65$. 
In this case we obtain $\theta_{\rm min}=-0.01\pi$, i.e., almost in-phase, 
and the strength distribution is similar to Fig.~\ref{fig:SDIM}. Thus the in-phase 
character of the low-lying mode which favors the overlap of the two kinds 
of particles is clearly seen in the determination of $F(\theta_{\rm min})$.

\subsubsection{dipole}

Dipole strength distributions for the in-phase and the out-of-phase 
operators are shown in Fig.~\ref{fig:DSD-IO}. The strong peak 
in the in-phase strength distribution corresponds to the center-of-mass 
oscillation of the whole system. For many-particle systems 
confined in a common oscillator potential 
the center-of-mass motion is decoupled from other (intrinsic) degrees 
of freedom of the system, 
resulting in the oscillation with the same  frequency $\omega_0$. 
This relation can be represented by the commutation relation of 
the dipole operator and the Hamiltonian, and should hold also 
within RPA \cite{Thouless}. 
In the present numerical calculation, however, 
the state at $\Omega=\omega_0$ does not exhaust the whole strength 
for the in-phase oscillation; the largest deviation occurs at 
$h/g=-1$ having 80\% of the EWS. 
This is because the single-particle model space in our numerical 
calculation is not sufficient to completely decouple the center of 
mass motion, especially at strong coupling cases where the deviation 
of the single particle potential from the oscillator becomes large. 
The $m_1$ sum rule is, nevertheless, almost satisfied as mentioned 
earlier, and is determined by the total number of bose and fermi 
particles, where the latter number is dependent on the value of $h/g$. 
Smaller value of the sum rule percentage at $h/g=-1$ is 
mainly due to the appearance of the almost degenerate 
state ($\Delta\omega/\omega_0\simeq 10^{-5}$) at this energy.

All other dipole oscillations should be orthogonal to the 
center-of-mass 
motion and thus are out-of-phase type in character. One may note 
that at  repulsive values of $h$ the dominant part of the 
out-of-phase strength is located below $\omega_0$, although 
the strength carried by each individual state is not very large. 
As discussed in Ref.~\cite{MSY}, this may reflect the 
ground state density distribution which favors the out-of-phase 
oscillation by making the overlap of boson and fermion 
distributions smaller. Figure \ref{fig:TDDO} shows 
transition densities for the out-of-phase type oscillations at 
$h/g=5.0,1.0,-3.0$ which carry the largest strengths; 
$\Omega/\omega_0=0.700$ for $h/g=5.0$, 
$0.956$ for $h/g=1.0$ and $1.14$ for $h/g=$-3.0.
Bosons and fermions move in the opposite directions 
so as to make the overlap smaller around the surface of the 
boson distribution.
One can observe that 
the bosonic transition density is large and robust, while the 
fermionic one is feeble and is spread over the whole system. This 
situation may be analogous to the soft dipole mode speculated in 
neutron-rich nuclei wherein the protons oscillate almost free in 
the sea of neutrons\cite{Ikeda}. Dynamical structure factors 
corresponding to these states are shown in Fig.~\ref{fig:DSF-D}. Here the 
bosonic and the fermionic responses are shown together with the 
one for a hypothetical out-of-phase type probe. By changing the 
momentum transfer one would find a structure corresponding to 
the oscillation of the two kinds of particles. 

It was suggested in \cite{Molmer,Minguzzi} that for a very strong repulsion 
between bosons and fermions with sufficient number of particles, the 
system may become unstable towards a phase separation of the two 
kinds of particles. Out-of-phase dipole strength distribution in 
Fig.~\ref{fig:DSD-IO} indeed show a softening of the strength distribution 
at large value of $h/g$. We cannot conclude from the present calculation, 
however, if this tendency is related to the mentioned instability.

\subsubsection{quadrupole}

One finds from the quadrupole strength distribution in Fig.~\ref{fig:QSD-FB} that 
the fermionic strength is split into low- and high-energy parts, while 
the bosonic one is concentrated. The higher strengths for the fermions 
are due to the 2$\hbar\omega_0$ excitation of the particle from the 
occupied single-particle states and are similar in character to the 
bosonic excitation. In contrast, the lower fermion strengths come 
from the matrix elements of the quadrupole operator $F_2$ which 
re-orient the single particle states within the same oscillator 
shell of $N=2n+\ell$ around the Fermi surface. This transition normally 
involves a change of nodes by one, see Fig.~\ref{fig:FSPE}, and the corresponding 
strengths are smaller than the higher ones. One may note that the quadrupole 
bosonic strength is located slightly higher than $\sqrt{2}\omega_0$ as 
expected for the  collective oscillation in the large $N$ limit\cite{Dalfovo}, 
which may be due to a rather small number of particles in the present 
calculation. 

We note that the strength distribution tends to be fragmented at large 
(repulsive) value of the interaction parameter $h/g$ as seen in Fig.~\ref{fig:QSD-FB}. 
This is because the average potential 
deviates appreciably from the harmonic oscillator potential, giving rise 
to a dispersion in the fermion particle-hole energies, see Fig.~\ref{fig:FSPE}. Strong 
fermion-boson interaction would then be responsible to scatter the 
bosonic strengths, too.

\subsubsection{octupole}

Octupole strength distribution is shown in Fig.~\ref{fig:OSD-FB}. Here again the 
fermionic strength is split into $\hbar\omega_0$ and 
$3\hbar\omega_0$ regions due to the character of 
the multipole operator $F_3$. We note also that the frequency of 
the bosonic oscillation is much larger than the hydrodynamic 
value $\sqrt{3}\omega_0$\cite{sum}. 

We find a striking change in the fermionic strength distributions 
depending 
on the sign of the bose-fermi interaction. While for a strongly 
repulsive case ($h/g=5.0$) the region between the low- and high-lying 
states is almost filled up by small strengths, the one for a strongly 
attractive case ($h/g=-3.0$) shows a large gap in the strength 
distribution. This may be traced back to the structure of the single-particle 
states in Fig.~\ref{fig:FSPE}. The large strengths around $3\omega_0$ are due to 
high orbital angular momentum states with $\ell_p-\ell_h=3$ which stay 
robust against $h/g$ as discussed earlier. Smaller strengths due to 
low $\ell_p$ and $\ell_h$ states are, on the other hand, sensitive to 
the interaction, and the correponding particle-hole energies are 
below (for $h/g>0$) or above (for $h/g<0$) the unperturbed value 
$3\omega_0$. This may be understood by noting that the particles 
passing through the center (small $\ell$ orbit) feel the shallow 
(for $h/g>0$) or deep (for $h/g<0$) potentials and are thus 
more easy or hard to excite. For the strengths around $\omega_0$ 
different combinations of orbits are involved, and the strengths 
are broadened by the deviation  from the oscillator potential.

\subsection{Time evolution}

To study the collective oscillation of the cold atomic gases, the 
current experiments on BEC exert a time-dependent field on the system 
and then observe the time development of the shape and the size 
of the condensate\cite{bec}.  This procedure generally excites a number 
of normal modes, and one may in principle be able to resolve each mode 
by performing 
Fourier transform as far as the time duration is long enough. To 
simulate the situation we consider a time evolution of the system 
after one applies an external weak pulse of the step-function type, i.e., 
$\Delta V \propto F_L^\tau$ for $-\Delta t \leq t \leq 0$. The calculation is 
performed within the lowest order perturbation theory. 
($\Delta t$ is chosen as $\Delta t=10^{-2}\omega_0^{-1}$.)

\subsubsection{monopole}

First we consider a perturbation of  monopole type
\begin{equation}
    \Delta V=m\omega_0\Delta\omega\,\sqrt{4\pi}F_0^\tau
    \label{Tmono}
    \end{equation}
with $\Delta \omega\ll\omega_0$ which is applied to the ground state of the 
system for a short period $\Delta t$. The effective frequency of 
the external oscillator potential for fermions and/or  bosons is then 
changed into $\omega_0\pm\Delta\omega$ depending on the choice of 
$\tau$. Time dependence of the rms radius of bosons and fermions is plotted in Fig.~\ref{fig:TDO-M} 
at $h/g=-3$ for the fermionic external field, $\tau=f$. 
Frequency of the short period oscillation is related to 
the average frequency of the strength distribution, e.g., 
$\omega_{av}=1.99\omega_0$.
That of the long period one, on the other hand, 
reflects the width of the distribution. 
For instance, the slow decrease of the envelope 
for $\sqrt{\langle r_f^2\rangle}$ would imply 
a very narrow width of the strength distribution. 
The bottom of Fig.~\ref{fig:TDO-M} shows the Fourier transform of 
$\langle r_f^2\rangle(t)-\langle r_f^2\rangle_0$ 
which recovers the sharp peak structure of the strength distribution 
in Fig~\ref{fig:MSD-FB}. 
Time evolution of the bosonic radius shows a rather regular modulation, 
and its beat frequency is estimated about $0.125\omega_0$, 
which is consistent with the half width $\sigma/2=0.127\omega_0$. 

\subsubsection{dipole}

For the dipole case we take the perturbing potential
\begin{equation}
    \Delta V=-m\omega_0^2\Delta z\,\sqrt{\frac{4\pi}{3}}F_1^\tau,
    \label{Tdip}
    \end{equation}
where $\Delta z(\ll \xi)$ measures the shift of the centers of the 
oscillator potentials felt by fermions and/or bosons. In Fig.~\ref{fig:TDO-D} we show the 
time dependence of the center-of-masses of bosons and 
fermions for the out-of-phase external dipole impulse, $\tau=-$, at $h/g=5$. 
The frequency of the oscillation is consistent with
the average frequency $0.90\omega_0$ for the dipole strength distribution. 
The amplitude of the oscillation shows an irregular behavior, 
suggesting that the strength distribution does not 
have a simple structure.
The lowest part of the 
figure shows the  Fourier transform of the relative distance of the 
fermion and boson center-of-mass positions, 
$N_f \langle z_f \rangle -N_b \langle z_b \rangle$, 
which is sufficient to recover the behavior of the original 
strength distribution given in Fig.~\ref{fig:DSD-IO}.

\section{Summary and Conclusion}

In the present paper we performed an RPA calculation of the polarized 
bose-fermi mixed system of alkali-metal gases at zero temperature. We solved 
a coupled Gross-Pitaevskii-Thomas-Fermi equations  to obtain the ground 
state for several values of the boson-fermion interaction strength 
with a fixed value of the boson-boson interaction. Single particle 
states of bose and 
fermi particles are calculated based on the mean field produced by 
the obtained ground state density. The density distribution constructed 
from the occupied single-particle orbits agrees well with the original 
one, suggesting the self-consistency of the calculation. 
We calculated and diagonalized the RPA matrix to obtain excitation 
energies and wave functions for  multipoles $L=0,1,2,3$. We then calculated 
strength distributions for bosonic/fermionic and in-phase/out-of-phase 
multipole operators, and also transition densities and dynamical structure 
factors for some of the collective states. 

We first calculated energy weighted moments of the strengths  to study 
average behavior of the strength distribution. Comparison of the 
first moments with the 
sum rule predictions for $L=0,1,2$ shows that the RPA configuration 
space is sufficient to within 1\%. A glance at the monopole distribution 
suggests that fermions and bosons are moving rather independently for the 
adopted parameters , which turns out not to be 
the case if we study the dynamical structure factors. We find, in fact, 
that the long wavelength oscillation of fermions in the surface is 
coupled to the internal short-wavelength oscillation of bosons. For 
a strong attractive boson-fermion interaction, the calculation suggests 
a softening of in-phase monopole oscillation. We studied also the structure 
of the low-lying out-of-phase type dipole modes. Lowering of the 
energy is related to the ground state density distribution of bosons and 
fermions which favors the out-of-phase oscillation. Transition densities 
for these modes suggest that the bosons oscillate on their way through the 
cloud of fermions, which is analogous to the soft dipole mode discussed 
in neutron-rich nuclei. We also calculated the strength distributions for 
quadrupole and octupole modes, and studied, in particular, the 
origin of the fragmentation of fermionic strengths. Finally we considered 
time dependent behavior of the trapped bose-fermi system after an impulse 
of a multipole external field. By Fourier transforming the time-dependent 
oscillating behavior of the system, we could recover the gross structure 
of the strength distribution.



\begin{figure}
\caption{\label{fig:FSPE}Fermion single particle energies for $h/g=5.0,1.0,-3.0$ 
against orbital angular momentum quantum number. Energies are 
measured from the Fermi energy. Eigenstates which are degenerate 
at $h/g=0$ are connected by solid lines. }
\end{figure}
\begin{figure}
\caption{\label{fig:AFCE}Average frequencies of the collective excitations in the 
sum rule formalism. Average frequencies $\bar\omega$, Eq.~(\ref{average}),  
in units of $\omega_0$ is plotted against interaction strength ratio
$h/g$ with fixed $g$. Solid and dotted lines respectively show the 
in- and out-of-phase oscillations. For a comparison, average 
frequencies calculated from the ground state expectation values of 
double commutators , see Ref.~\protect \cite{MSY}, 
are plotted by dashed (in-phase) and dot-dashed (out-of-phase) lines.}
\end{figure}
\begin{figure}
\caption{\label{fig:MSD-FB}Monopole strength distribution for fermionic and bosonic 
operators, $F_0^f$ and $F_0^b$, at $h/g=$5.0, 1.0 and -3.0. Strengths 
in units of $\xi^4$ are plotted against excitation energy. }
\end{figure}
\begin{figure}
\caption{\label{fig:DSF-M}Dynamical structure factors for the states in which in-phase 
monopole strengths are the largest. These states correspond to the 
ones at $\Omega/\omega_0=2.03$ for $h/g=5.0$, 
at $2.01$ for $h/g=1.0$ and at $1.98$ for 
$h/g=$-3.0. Dotted and dashed lines correspond to the fermion and boson 
transition densities, $\delta\rho_{\nu 0}^{f,b}$, while solid lines are for the in-phase one, 
$\delta\rho_{\nu 0}^+= \delta\rho_{\nu 0}^f + \delta\rho_{\nu 0}^b$. 
The abscissa 
is $q\xi/\hbar$, where $q$ denotes the momentum transferred to the system 
by external probes.}
\end{figure}
\begin{figure}
\caption{\label{fig:SDIM}Strength distribution for the in-phase monopole operator 
$F_0^+$ at $h/g=$-6.65 plotted against $\Omega/\omega_0$.}
\end{figure}
\begin{figure}
\caption{\label{fig:SD-theta}Strength distribution for the $\theta$-dependent 
monopole operator $F_0(\theta)$ at $h/g=5.0$. Upper figure shows the 
strength distribution for the operator 
$F(\theta_{\rm min}=-0.12\pi)$ which was determined so as 
to minimize the average frequencey $\bar\omega$. 
Middle figure is the one for the operator $F_\perp=
F(\theta=0.38\pi)$ which is perpendicular to the above operator. 
Lower figure is the strength distribution for the operator 
$F(\theta_{\rm max}=0.27\pi)$ which was determined to maximize 
the average energy. }
\end{figure}
\begin{figure}
\caption{\label{fig:DSD-IO}Dipole strength distribution for in- and out-of-phase 
operators, $F_1^+$ and $F_1^-$. Strengths are measured in 
units of $\xi^2$.}
\end{figure}
\begin{figure}
\caption{\label{fig:TDDO}Transition densities for the dipole oscillations with 
largest strengths of the out-of-phase type. Excitation energies 
of these states are $\Omega/\omega_0=$0.700 for $h/g=$5.0,
0.956 for $h/g=$1.0 and 1.14 for $h/g=$-3.0. 
Solid and dotted lines 
respectively show the fermionic and bosonic transition densities, 
$\delta\rho_{\nu 1}^{f,b}$, in units of $\xi^3$.}
\end{figure}
\begin{figure}
\caption{\label{fig:DSF-D}Dynamical structure factors for the same states as given in 
Fig.~\ref{fig:TDDO}. Dotted and dashed lines are those for the fermionic and 
bosonic transition densities while solid lines are those for the out-of-phase one, 
$\delta\rho_{\nu 1}^- = \delta\rho_{\nu 1}^f - \delta\rho_{\nu 1}^b$. }
\end{figure}
\begin{figure}
\caption{\label{fig:QSD-FB}Quadrupole strength distributions for fermionic and bosonic 
operators, $F_2^f$ and $F_2^b$, at $h/g=$5.0, 1.0 and -3.0. }
\end{figure}
\begin{figure}
\caption{\label{fig:OSD-FB}Octupole strength distributions for fermionic and bosonic 
operators, $F_3^f$ and $F_3^b$, at $h/g=$5.0, 1.0 and -3.0. }
\end{figure}
\begin{figure}
\caption{\label{fig:TDO-M}Time dependent oscillation of the bose-fermi system 
at $h/g=$-3.0 after an 
external impulse of fermionic monopole-type, Eq.~(\ref{Tmono}). 
Time dependences of the root-mean-square radii (in units of $\xi$) 
of fermions (upper figure) and of bosons (middle figure) are shown. 
Parameters of the impulse are: $\Delta\omega/\omega_0=0.01$ and 
$\omega_0\Delta t=0.01$. Differential equation in time is solved with 
the time-mesh of 0.05$\omega_0^{-1}$ up to $t_{\rm max}=100/\omega_0$. 
The lowest
figure shows the Fourier transform (in an arbitrary unit) of the 
fermionic rms radius deviation from the ground state, 
$\langle r_f^2 \rangle(t)- \langle r_f^2 \rangle_0$,
which corresponds to the fermionic strength distribution given in Fig.~\ref{fig:MSD-FB} 
at $h/g=-3.0$.}
\end{figure}
\begin{figure}
\caption{\label{fig:TDO-D}Time dependence of the oscillations of fermions and 
bosons at $h/g=5.0$ for an out-of-phase type external dipole impulse, 
Eq.~(\ref{Tdip}). Upper and middle figures show 
the fermion and boson center-of-mass in the unit of $\xi$ against 
elapsed time in the unit of $\omega_0^{-1}$. Parameters are given 
by $\Delta z/\xi=0.01$ and $\omega_0\Delta t=0.01$. Calculation is 
performed with time-mesh of 0.05$\omega_0^{-1}$ and up 
to $t_{\rm max}=100/\omega_0$. The 
lowest figure shows the Fourier transform (in an arbitrary unit) of the 
relative distance of 
fermions and bosons,
$N_f \langle z_f \rangle -N_b \langle z_b \rangle $, 
with an energy resolution of $\sim 10^{-2}$. 
The figure corresponds to the strength distribution 
of the out-of-phase type in Fig.~\ref{fig:DSD-IO} at $h/g=5.0$.}
\end{figure}
%
\end{document}